\documentstyle[12pt,a4]{article}
\author{Afraimovich~E.~L., Kosogorov~E.~A.,
        Perevalova~N.~P.,  \and and Plotnikov~A.~V.     \\
        Institute of Solar-Terrestrial Physics SD RAS,\\
        p.~o.~box~4026, Irkutsk, 664033, Russia\\
        fax: +7 3952 462557; e-mail:~afra@iszf.irk.ru}

\title{The use of GPS-arrays in detecting shock-acoustic waves
       generated during rocket launchings}

\date{}

\begin{document}
\sloppy
\maketitle
\begin{abstract}
This paper is concerned with the form and dynamics of
shock-acoustic waves (SAW) generated during rocket launchings. We
have developed a method for determining SAW parameters (including
angular characteristics of the wave vector, and the SAW phase
velocity, as well as the direction towards the source) using
GPS-arrays whose elements can be chosen out of a large set of
GPS-stations of the global GPS network. Contrary to the
conventional radio-probing tecniques, the proposed method
provides an estimate of SAW parameters without a priori
information about the site and time of a rocket launching. The
application of the method is illustrated by a case study of
ionospheric effects from launchings of launch vehicles (LV)
Proton and Space Shuttle from space-launch complexes Baikonur
and Kennedy Space Center (KSC) in 1998 and 1999 (a total of five
launchings). The study revealed that, in spite of a difference of
LV characteristics, the ionospheric response for all launchings
had the character of an $N$--wave corresponding to the form of a
shock wave, regardless of the disturbance source (rocket
launchings, industrial explosions). The SAW period $T$ is 270--360~s,
and the amplitude exceeds the standard deviation of TEC
background fluctuations in this range of periods under quiet and
moderate geomagnetic conditions by factors of 2 to 5 as a
minimum. The angle of elevation of the SAW wave vector varies
from 30${}^\circ$ to 60${}^\circ$ , and the SAW phase velocity
(900--1200~m/s) approaches the sound velocity at heights of the
ionospheric $F$-region maximum. The position of the SAW source,
inferred by neglecting refraction corrections, corresponds to the
segment of the LV Proton path at a distance no less than
700--900~km from the launch pad, and to the LV flying altitude no
less than 100~km. For the Space Shuttle LV the position of the
SAW source corresponds to the segment of the trajectory at a
distance of at least 200--500~km from the launch pad and to the LV
flight altitude of at least 100~km. Our data are consistent with
the existing view that SAW are generated during a nearly
horizontal flight of the rocket with its engine in operation in
the acceleration segment of the path at 100--130~km altitudes in
the lower atmosphere.
\end{abstract}

\section{Introduction}
\label{LAUNCH-sect-1}

A large number of publications, including a variety of thorough
reviews (Karlov et al., 1980; Mendillo, 1981; 1982; Nagorsky,
1998; 1999) are devoted to the study of the ionospheric response
of the shock wave produced during launchings of powerful rockets,
among them launch vehicles (LV). Scientific interest in this
problem is due to the fact that such launchings can be regarded
as active experiments in the Earth's atmosphere, and they can be
used in solving a wide variety of problems in the physics of the
ionosphere and radio wave propagation.

These investigations have also important practical implications
since they furnish a means of substantiating reliable signal
indications of technogenic effects (among them, unauthorized),
which is necessary for the construction of an effective global
radiophysical subsystem for detection and localization of these
effects. Essentially, existing global systems of such a purpose
use different processing techniques for infrasound and seismic
signals. However, in connection with the expansion of the
geography and of the types of technogenic impact on the
environment, very challenging problems heretofore have been those
of improve the sensitivity of detection and the reliability of
measured parameters of the sources of impacts, based also on
using independent measurements of the entire spectrum of signals
generated during such effects.

To solve the above problems requires reliable information about
fundamental parameters of the ionospheric response of the shock
wave, such as the amplitude and the form, the period, the phase
and group velocity of the wavetrain, as well as angular
characteristics of the wave vector. Note that for the designating
the ionospheric response of the shock wave, the literature uses
terminology incorporating a different physical interpretation,
among them the term `shock-acoustic wave' (SAW)~--- Nagorsky
(1998). For convenience, in this paper we shall use this term
despite the fact that it does not reflect essentially the
physical nature of the phenomenon.

Published data on fundamental parameters of the SAW differ
greatly. According to a review by Karlov et al. (1980), the
oscillation period of the ionospheric response of the SAW,
recorded during launchings of the Apollo mission rockets,
varied from 6 to 90~min, and the propagation velocity was in the
range from 600 to 1670~m/s. The added complication is also that
following the first SAW response arriving at the detection point,
a `second' wave propagating with the sound velocity at the ground
level is often recorded (Karlov et al., 1980; Calais and Minster,
1996). A `third' wave with large periods and subsonic propagation
velocities is occasionally also detected (Nagorsky, 1998; 1999).
This wave is identified as an exciting Acoustic-Gravity Wave
(AGW) during a rocket launching.

There is an even greater uncertainty in the localization of the
region of SAW generation, which is necessary both for an
understanding of the physical mechanisms of the phenomenon, and
for a scientific justification of the solution of applied
problems. Published data are discussed in more detail in
Section~5, by comparison with our obtained results.

The lack of comprehensive, reliable data on SAW parameters is due
primarily to the limitations of existing experimental methods and
detection facilities. The main body of data was obtained by
measuring the frequency Doppler shift at vertical and
oblique-incidence ionospheric soundings in the HF range (Jacobson
and Carlos, 1994; Nagorsky, 1998). In some instances the
sensitivity of this method is sufficient to detect the SAW
reliably; however, difficulties emerge when localizing the region
where the detected signal is generated. These problems are caused
by the multiple-hop character of HF signal propagation. This
gives no way of deriving reliable information about the phase and
group velocities of SAW propagation, estimating angular
characteristics of the wave vector, and, further still, of
localizing the SAW source.

Using the method of transionospheric sounding with VHF radio
signals from geostationary satellites, a number of experimental
data on SAW parameters were obtained in measurements of the
Faraday rotation of the plane of signal polarization which is
proportional to a total electron content (TEC) along the line
connecting the satellite-borne transmitter with the receiver
(Mendillo, 1981; 1982; Li et al., 1994). A serious limitation of
methods based on analyzing signals from geostationary satellites
is that their number is too small and ever decreasing with time,
and the distribution in longitude is non-uniform; this is
especially true in regard to the eastern hemisphere. Published
data from incoherent scatter (IS) radars are also scarce; as far
as our knowledge goes, there is only one publication (Noble,
1990) reporting the detection of long-period (gravity) waves over
the Arecibo IS station, following the launching of the Space
Shuttle LV on October~18, 1993.

A common limitation of the above-mentioned methods when
determining the SAW phase velocity is the necessity of knowing
the time of a rocket launching since this velocity is inferred
from the SAW delay with respect to the launch time assuming that
the velocity is constant along the propagation path, which is
quite contrary to fact.

For determining the above-mentioned reasonably complete set of
SAW parameters, it is necessary to have appropriate spatial and
temporal resolution which cannot be ensured by existing, very
sparse, networks of ionozondes, oblique-incidence radio sounding
paths, and IS radars. Furthermore, the creation of these
facilities involves developing dedicated equipment, including
powerful radio transmitters which contaminate the radio medium.
As a result, the above systems cannot serve as an efficient basis
for a global radiophysical subsystem for detection and
localization of technogenic effects which must provide a
continuous, global monitoring.

The advent and evolution of a Global Positioning System, GPS, and
also the creation on its basis of widely branched networks of GPS
stations (at least 600 sites at the end of 1999, the data from
which are placed on the INTERNET) opened up a new era in remote
ionospheric sensing. In the immediate future this network will
undergo a substantial expansion by integration with the GLONASS
positioning system (Klobuchar, 1997). In addition, researchers
have also access to a standard software which permits them to
unify a preprocessing of the data from multichannel two-frequency
GPS receivers. In this context, a major effort must go into
experimental and theoretical explorations of the possibilities of
detecting technogenic perturbations of the ionosphere, as well as
into the development of appropriate algorithms and software for a
reprocessing of the data from GPS networks.

Currently some authors have embarked on an intense development of
methods for detecting the ionospheric response of strong
earthquakes (Calais and Minster, 1995), rocket launchings (Calais
and Minster, 1996), and industrial surface explosions
(Fitzgerald, 1997; Calais et al., 1998a; Calais et al., 1998b).
In the cited references the SAW phase velocity was determined by
the `crossing' method by estimating the time delay of SAW arrival
at subionospheric points corresponding to different GPS
satellites observed at a given time. However, the accuracy of
such a method is rather low because the altitude at which the
subionospheric points are specified, is determined in a crude
way.

The goal of this paper is to develop a method for determining SAW
parameters (including the phase velocity, angular characteristics
of the SAW wave vector, the direction towards the source, and the
source location) using GPS-arrays whose elements can be chosen
out of a large set of GPS stations of a global IGS network.
Contrary to existing radio techniques, this method estimates SAW
parameters without a priori information about the location and
time of a rocket launching.

The emphasis in this paper is on the ionospheric response of the
SAW which is recorded as a `first' wave, although our developed
methods make it also possible to analyze other wave
manifestations of rocket launchings. This will be the subject of
future work.

Section~2 presents a description of the
geometry of experiments and general information about the LV
launchings from the Baikonur and Kennedy Space Center (KSC)
cosmodromes which are analyzed in this study. A brief account of
the proposed method is presented in Section~3.
Measurements of SAW parameters at different GPS-arrays during
rocket launchings are discussed in Section~4.
Section~5 discusses experimental results and
compares them with the data reported by other authors. It is
beyond the scope of this paper to consider the physical
mechanisms of SAW generation and propagation.

\section{The geometry and general characterization of experiments}
\label{LAUNCH-sect-2}

This paper presents the results derived from determining the key
SAW parameters during ascents of the Proton and Space Shuttle
launch vehicles (LV) from cosmodromes Baikonur (45.6${}^\circ N$,
63.3${}^\circ E$) and KSC (28.5${}^\circ N$, 279.3${}^\circ E$)
in 1998 and 1999 (a total of five launchings).

In spite of the large number of GPS stations, the selection of
GPS-arrays for detecting SAW generated during rocket launchings is
made difficult by the fact that the LV flight paths for both
cosmodromes pass either over the Atlantic ocean (Space
Shuttle LV) or over sparsely populated regions of Kazakhstan
(Proton LV). Nevertheless, it was possible to utilize a sufficient
number of GPS stations in order for this method to be implemented.
Table~1 presents the geographic coordinates of
the GPS stations used as GPS-array elements.

Fig.~1 portrays the geometry of experiments
during Proton LV launchings from the Baikonur cosmodrome. The
dash line roughly corresponds to the horizontal projection of the
rocket flight trajectory with orbital inclination
$\psi=51.6{}^\circ$ (Nagorsky, 1999). The symbol $\boldmath
\diamond$ marks the pad site. Bold dots mark the position of GPS
stations, and upper-case lettering corresponds to their names.
The scaling of the coordinate axes is selected from
considerations of an approximate equality of the linear
dimensions along the latitude and longitude.

Below is a summary of results derived from detecting one standard
launching (on November~20, 1998) and two abortive launchings
(July~5 and October~27, 1999) of the Proton LV. The following
information about rocket launchings from the Baikonur Cosmodrome
as extracted from the INTERNET sites: http//www.flatoday.com;
http://www.spacelaunchnews.com; and http://www.ksc.nasa.gov.
Official data on abortive Proton LV launches on July~5 and
October~27, 1999 were taken from the English version of the press
release issued by the Khrunichev Center which is available on:
http://www.isllaunch.com.

A Proton launch vehicle failed during launch on July~5, 1999
(Day~186). Initial analyses indicated a malfunction of the
Proton's second stage. Liftoff took place at 13{:}32~UT from a
launch pad of the Baikonur Cosmodrome. First stage flight was
normal, with second stage separation and ignition on time.
Telemetry first reported anomalous data at approximately 280
seconds into the flight, with deviations from the planned
trajectory appearing on ground tracking systems by 330 seconds.
By 390 seconds the LV was 14~km below its planned trajectory.
Ground systems tracked the upper stages and payload to impact.
The Proton's third stage, with the upper stage and its Raduga
spacecraft payload, fell in the Karaganda region of Kazakhstan.

The Proton LV was launched from a launch pad at Baikonur
Cosmodrome on October~27 (Day~300), 1999 at 16{:}16~UT. Based on
preliminary data, failure of one of the main second stage engines
occurred 222 seconds after liftoff, followed by failure of all
engines of the stage. The second stage together with the upper
stage and satellite fell back to earth in the Karaganda region of
Kazakhstan.

In a typical Proton launch, the vehicle's six first-stage
engines ignite 1.6~s before liftoff. Stage two ignition occurs
approximately two minutes into flight, four seconds prior to
jettison of the first stage. Stage three vernier engine ignition
occurs at 330 seconds, with separation of the second and third
stages taking place 3.5~s later. The stage three main engine
ignition occurs 2.5~s after separation.

Unfortunately, we were unable to obtain more detailed information
about the LV launch schedule on the above days. However, for a
further analysis we need only know that these failures occurred
after the second stage was in operation longer than 200~s and
100~s for the July~5 and October~27, 1999 launches, respectively.

General information regarding these launches, as well as the
scheduled launch of November~20, 1998 (Day~324) is summarized in
Table~2 (including the start time $t_0$ of UT,
the start time of LT for one of the GPS-array sites, the orbital
inclination $\psi$ to the equatorial plane, and the level of
geomagnetic disturbance according to $Dst$-variation data). It
was found that the deviation of $Dst$ for the selected days was
quite moderate thus enabling the SAW to be identified.

Fig.~2 presents the experimental geometry during
Space Shuttle LV launchings from the KSC Cosmodrome (the
designations are the same as in Fig.~1). The
Space Shuttle Discovery spacecraft (STS-90) and Space Shuttle
Columbia (STS-95) were launched on April~17 (Day~107) and on
October~29 (Day~302), respectively. General information about
these launches is also summarized in Table~2.
The horizontal projection of the rocket trajectory for the
October~29 launching is constructed using the INTERNET data.
The level of geomagnetic disturbance as deduced from the
$Dst$-variation data was also found to be quite moderate for
these launches, which made it possible to reliably identify the
SAW.

\section{Methods of determining shock-acoustic wave characteristics
using GPS-arrays}
\label{LAUNCH-sect-3}

The standard GPS technology provides a means for wave
disturbances detecion based on phase measurements of TEC at each
of spaced two-frequency GPS receivers. A method of reconstructing
TEC variations from measurements of the ionosphere-induced
additional increment of the group and phase delay of the
satellite radio signal was detailed and validated in a series of
publications (Calais and Minster, 1995, 1996; Fitzgerald, 1997).
We reproduce here only the final formula for phase measurements

\begin{equation}
\label{LAUNCH-eq-1}
I=\frac{1}{40{.}308}\frac{f^2_1f^2_2}{f^2_1-f^2_2}
                           [(L_1\lambda_1-L_2\lambda_2)+const+nL]
\end{equation}
where $L_1\lambda_1$ and $L_2\lambda_2$ are additional paths of
the radio signal caused by the phase delay in the
ionosphere,~(m); $L_1$ and $L_2$ represent the number of phase
rotations at the frequencies $f_1$ and $f_2$; $\lambda_1$ and
$\lambda_2$ stand for the corresponding wavelengths,~(m);
$const$ is the unknown initial phase ambiguity,~(m); and $nL$~
are errors in determining the phase path,~(m).

Phase measurements in the GPS can be made with a high degree of
accuracy corresponding to the error of TEC determination of at
least $10^{14}$~м${}^{-2}$ when averaged on a 30-second time
interval, with some uncertainty of the initial value of TEC,
however (Hofmann-Wellenhof et al., 1992). This makes possible
detecting ionization irregularities and wave processes in the
ionosphere over a wide range of amplitudes (up to $10^{-4}$ of
the diurnal TEC variation) and periods (from 24 hours to 5 min).
The unit of TEC, which is equal to $10^{16}$~м${}^{-2}$~($TECU$)
and is commonly accepted in the literature, will be used in the
following.

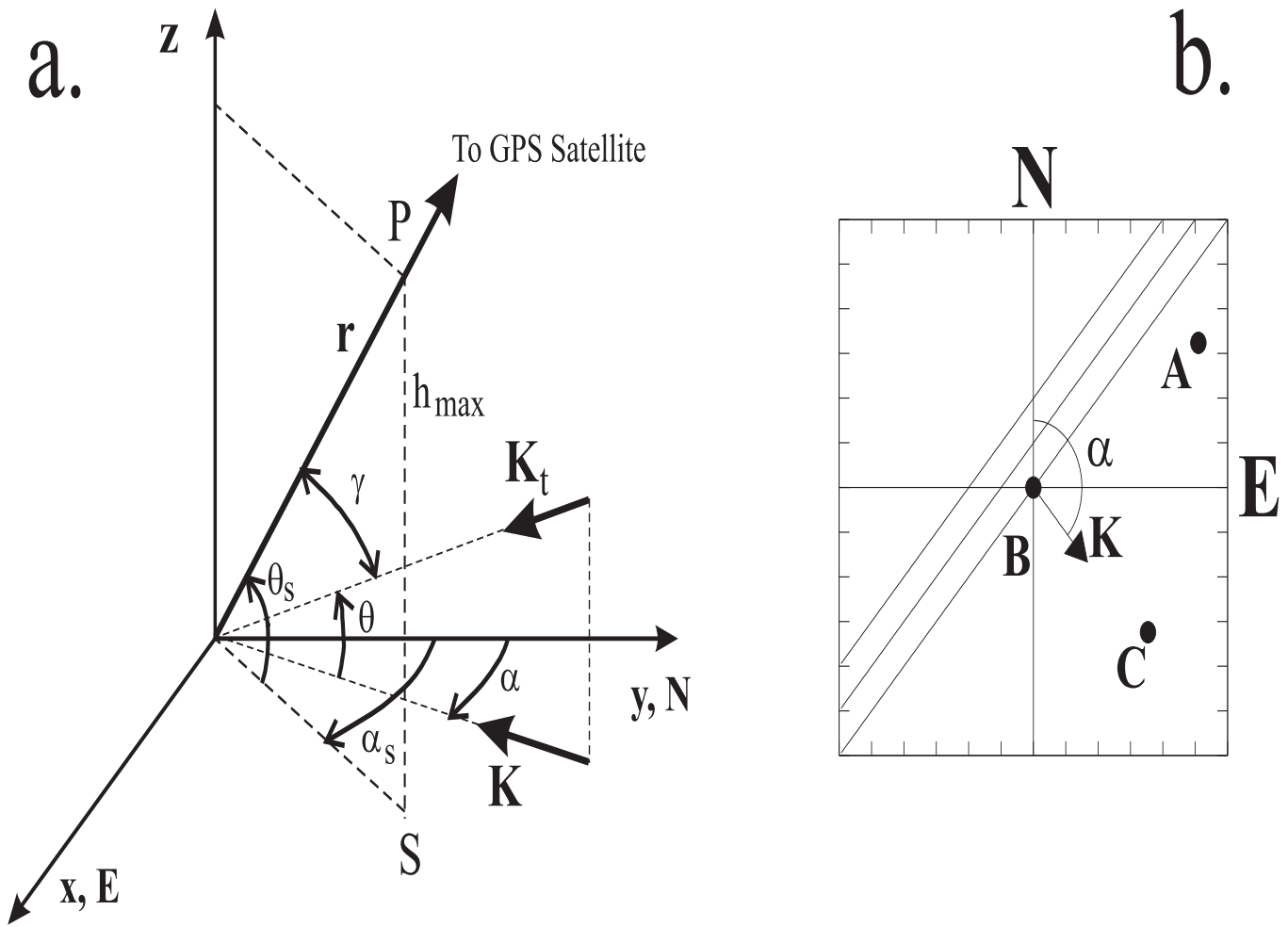
Fig.~3a gives a schematic representation of the
transionospheric sounding geometry. The axes $z$, $y$, and $x$ are
directed, respectively, zenithward, northward ($N$) and eastward
($E$). P~--- point of intersection of Line-of-Sight (LOS) to the
satellite with the maximum of the ionospheric $F2$-region; S~---
subionospheric point; $\alpha_s$~---the azimuthal angle, counted
off from the northward in a clockwise direction; and
$\theta_s$~---the angle of elevation between the direction
$\boldmath r$ along LOS and the terrestrial surface at the
reception site.

In some instances a convenient way of detecting and determining
the ionospheric response delay of the shock wave involves
inferring the frequency Doppler shift $F$ from TEC series
obtained by formula~(\ref{LAUNCH-eq-1}). Such an approach is also
useful in comparing TEC response characteristics from the GPS
data with those obtained by analyzing VHF signals from
geostationary satellites, as well as in detecting the shock wave
in the HF range. To an approximation sufficient for the purpose
of our investigation, a corresponding relationship was obtained
by K. Davies (1969)

\begin{equation}
\label{LAUNCH-eq-2}
F=13{.}5\times 10^{-8}I'_t/f
\end{equation}
where $I'_t$ stands for the time derivative of TEC. Relevant results
derived from analyzing the $F(t)$-variations calculated for the
`reduced' frequency of 136~MHz are discussed in
Section~4.

The correspondence of space-time phase characteristics, obtained
through transionospheric soundings, with local characteristics of
disturbances in the ionosphere was considered in detail in a wide
variety of publications (Bertel et al., 1976; Afraimovich et al.,
1992; Mercier and Jacobson, 1997) and is not analyzed at
length in this study. The most important conclusion of the cited
references is the fact that, as for the extensively exploited
model of a `plane phase screen' disturbances $\Delta I(x,y,t)$ of
TEC faithfully copy the horizontal part of the corresponding
disturbance $\Delta N(x,y,z,t)$ of local concentration,
independently of the angular position of the source, and can be
used in experiments on measuring the wave disturbances of TEC.

However, the TEC response amplitude experiences a strong azimuthal
dependence caused by the integral character of a transionospheric
sounding. As a first approximation, the transionospheric sounding
method is responsive only to TIDs with the wave vector $\boldmath
K_t$ perpendicular to the direction $\boldmath r$. A corresponding
condition for elevation $\theta$ and azimuth $\alpha$ of an arbitrary
wave vector $\boldmath K_t$ normal to the direction $\boldmath r$,
has the form

\begin{equation}
\label{LAUNCH-eq-3}
\theta=\arctan(-\cos(\alpha_s-\alpha)/\tan\theta_s)
\end{equation}

We used formula~(\ref{LAUNCH-eq-3}) in determining the elevation
$\theta$ of $\boldmath K_t$ from the known mean value of azimuth
$\alpha$ by Afraimovich et al. (1998) -- see Sections
~3.2 and~4.

\subsection{Detection and determination of the horizontal phase
velocity $V_h$ and the direction $\alpha$ of the SAW phase front
along the ground by GPS-arrays}
\label{LAUNCH-sect-3.1}

In the simplest form, space-time variations of the TEC $\Delta
I(t,x,y)$ in the ionosphere, at each given time $t$ can be
represented in terms of the phase interference pattern that moves
without a change in its shape (the solitary, plane travelling
wave)

\begin{equation}
\label{LAUNCH-eq-4}
\Delta I(t,x,y)=\delta\sin(\Omega t-K_xx-K_yy+\varphi_0)
\end{equation}
where $\delta$, $K_x$, $K_y$, $\Omega$~--- are the amplitude,
the x- and y-projections of the wave vector {\boldmath $K$}, and
the angular frequency of the disturbance, respectively;
$T=2\pi/\Omega$, $\Lambda=2\pi/|K|$ is its period and wavelength;
and $\varphi_0$~ is the initial phase of the disturbance. The
vector $\boldmath K$ является is a horizontal projection of the
full vector $\boldmath K_t$ (Fig.~3a).

At this point, it is assumed that in the case of small spatial
and temporal increments (the distances between GPS-array sites
are less than the typical spatial scale of TEC variation, and the
time interval between counts is less than the corresponding time
scale), the influence of second derivatives can be neglected. The
following choices of GPS-arrays all meet these requirements.

For the SAW which have mostly the character of an $N$-wave (Li et
al., 1994; Calais and Minster, 1995, 1996; Fitzgerald, 1997;
Calais et al., 1998), the amplitude $\delta$
in~(\ref{LAUNCH-eq-4}) can be specified as (with an appropriate
choice of the model parameters)

\begin{equation}
\label{LAUNCH-eq-5}
\delta(t)=\exp[-(\frac{t-t_{max}}{t_d})^2]
\end{equation}
where $t_{max}$ is the time when the disturbance has a maximum
amplitude, and $t_d$ is the half-thickness of the `wave packet'.
We employed this model in simulating the SAW detected during
Proton LV launch on July~5, 1999 (at the left of
Fig.~4, and at the right of Fig.~5).

We now summarize briefly the sequence of data processing
procedures. Out of a large number of GPS stations, three sites
(A,B,C) are selected, the distances between which do not exceed
about one-half the expected wavelength $\Lambda$ of the
perturbation. Site B is taken to be the center of a topocentric
reference frame whose axis $x$ is directed eastward, and the axis
$y$ is directed northward. The receivers in this frame of
reference have the coordinates А($x_A$, $y_A$), В(0,0), С($x_C$,
$y_C$). Parallel lines in Fig.~3b are lines of
equal TEC; the arrow indicates the direction $\alpha$ of a
normal to these lines, that is, the direction of the wave vector
$\boldmath K$. Such a configuration of the GPS receivers
represents the GPS-array with a minimum number of the required
elements. In regions with a dense network of GPS sites, we can
obtain a large variety of GPS-arrays of a different configuration
enabling the acquired data to be checked for reliability; in this
paper we have exploited such a possibility.

The input data include series of the `oblique' value of TEC
$I_A(t)$, $I_B(t)$, $I_С(t)$, as well as corresponding series of
values of the elevation $\theta_s(t)$ and the azimuth
$\alpha_s(t)$ of the beam to the satellite calculated using our
developed CONVTEC software program which converts the
RINEX-files, standard for the GPS system, from the INTERNET
(Gurtner, 1993). For determining SAW characteristics, continuous
series of measurements of $I_A(t)$, $I_B(t)$, $I_C(t)$ are
selected with a length of at least a one-hour interval which
includes the start time.

To eliminate variations of the regular ionosphere, as well as
trends introduced by orbital motion of the satellite, a procedure
is used to remove the linear trend involving a preliminary
smoothing of the initial series with the selected time window.
This procedure is best suited to the detection of the signal such
as a single pulse ($N$-wave) when compared to the frequently used
band-pass filter (Li et al., 1994; Calais and Minster, 1995,
1996; Fitzgerald, 1997; Calais et al., 1998). A limitation of the
band-pass filter is the delay and the oscillatory character of
the response which gives no way of reconstructing the form of the
$N$-wave.

Series of the values of the elevation $\theta_s(t)$ and azimuth
$\alpha_s(t)$ of the beam to the satellite are used to determine
the location of the subionospheric point~S (see
Fig.~3a), as well as to calculate the elevation
$\theta$ of the wave vector $\boldmath K_t$ of the disturbance
from the known azimuth $\alpha$ (see formula~(\ref{LAUNCH-eq-3})
and Section~4.2).

The most reliable results from the determination of SAW
parameters correspond to high values of elevations $\theta_s(t)$
of the beam to the satellite because sphericity effects become
reasonably small. In addition, there is no need to recalculate
the `oblique' value of TEC $\Delta I(t)$ to the `vertical' value.
In this paper, all results were obtained for elevations
$\theta_s(t)$ larger than 30{}$^\circ$.

Since the distance between GPS-array elements (from several tens
of kilometers to a few thousand kilometers) is much smaller than
that to the GPS satellite (over 20000~km), the array geometry at
the height $h_{max}$ is identical to that on the ground.

Fig.~4a shows typical time dependencies of an
`oblique' TEC $\Delta I(t)$ at the GPS-array CHUM station near
the Baikonur Cosmodrome for the start day, July~5, 1999 (heavy
curve), one day before and after the start (thin lines). For the
same days; panel b shows TEC variations $\Delta I(t)$ with the
removed linear trend and a smoothing with the 5-min time window.
Variations in frequency Doppler shift $F(t)$; `reduced' to the
sounding signal frequency of 136~MHz, for three sites of the
array (SELE, CHUM, SHAS) for the launch day, July~5, 1999, are
presented in panel c. Day numbers, GPS station names and GPS PRN
satellite numbers are indicated in all panels. The arrows at the
abscissa axis indicate the start time $t_0$.

It is evident from Fig.~4 that fast $N$-shaped
oscillations with a typical period $T$ of about 300~s are clearly
distinguished among slow TEC variations. The oscillation
amplitude (up to 0{.}5~$TECU$) is far in excess of the TEC
fluctuation intensity during `background' days. Variations in
frequency Doppler shift $F(t)$ for spatially separated sites
(SELE, CHUM, SHAS) are well correlated but are shifted relative
to each other by an amount well below the period, which permits
the SAW propagation velocity to be unambiguously determined. A
time resolution of 30~s used in our study, which is standard for
the GPS data, is not quite sufficient for determining small
shifts of such signals with an adequate accuracy for different
sites of the array (Fig.~4c). Therefore, we used
a parabolic approximation of the $F(t)$-oscillations in the
neighborhood of minimum $F(t)$, which is quite acceptable when
the signal/noise ratio is high. In addition, we used different
combinations for three sites of the possible number of GPS-array
sites in the area of Baikonur and KSC Cosmodromes. Relevant
results for all combinations are presented in
Table~3.

With proper account of a good signal/noise ratio (larger than~1),
we determine the horizontal projection of the phase velocity
$V_h$ with the known coordinates of array sites A,~B,~C from
time $t_p$ shifts of a maximum deviation of the frequency Doppler
shift $F(t)$. Preliminarily measured shifts are subjected to a
linear transformation with the purpose of calculating shifts for
sites spaced relative to the central site northward $N$ and
eastward $E$. This is followed by a calculation of the $E$- and
$N$-components of $V_x$ and $V_y$, as well as the direction
$\alpha$ in the range of angles 0${}^\circ$--360${}^\circ$ and
the modulus $V_h$ of the horizontal component of the SAW phase
velocity

\begin{equation}
\label{LAUNCH-eq-6}
\begin{array}{rl}
\alpha&=\arctan(V_y/V_x)\\ V_h&=|V_xV_y|(V_x^2+V_y^2)^{-1/2}
\end{array}
\end{equation}
where $V_y$, $V_x$ are the velocities with which the phase front
crosses the axes $x$ and $y$. The orientation $\alpha$ of the
wave vector $\boldmath K$, which is coincident with the
propagation azimuth of the SAW phase front, is calculated
unambiguously in the range 0${}^\circ$--360${}^\circ$ subject to
the condition that arctan$(V_y/V_x)$ is calculated having regard
to the sign of the numerator and denominator.

The above method for determining the SAW phase velocity neglects
the correction for orbital motion of the satellite because the
estimates of $V_h$ obtained below exceed an order of magnitude as
a minimum the velocity of the subionospheric point at the height
$h_{max}$ for elevations $\theta_s>30^\circ$~ (Afraimovich et
al., 1998).

From the delay $\Delta t=t_p-t_0$ and the known path length
between the launch pad and the subionospheric point we calculated
also the SAW mean velocity $V_a$ in order to compare our obtained
estimates of the SAW phase velocity with the usually used method
of measuring this quantity.

\subsection{Determination of the elevation of the wave vector $\theta$
and the velocity modulus $V_t$ of the shock wave}
\label{LAUNCH-sect-3.2}

Afraimovich et al. (1992) showed that for an exponential
ionization distribution the TEC disturbance amplitude ($M$) is
determined by the aspect angle $\gamma$ between the vectors
$\boldmath K_t$ and $\boldmath r$ (see Fig.~3a),
as well as by the ratio of the wavelength of the disturbance
$\Lambda$ to the half-thickness of the ionization maximum $h_d$

\begin{equation}
\label{LAUNCH-eq-7}
M\propto\exp\left(-\frac{\pi^2h_d^2\cos^2\gamma}
                        {\Lambda^2\cos^2\theta_s}\right)
\end{equation}
where $\theta_s$ is the elevation of the wave vector $\boldmath r$.

In the case under consideration (see below), for the phase
velocity of order 1~km/s and for the period of about 200~s, the
wavelength $\Lambda$ is comparable with the half-thickness of the
ionization maximum $h_d$. When the elevations $\theta_s$ are
30{}$^\circ$, 45{}$^\circ$, 60{}$^\circ$ , the `beam-width' $M(\gamma)$
at 0{.}5 level is, respectively, 25{}$^\circ$, 22{}$^\circ$ and
15{}$^\circ$. If $h_d$ is twice as large as the wavelength, then
the beam tapers to 14{}$^\circ$, 10{}$^\circ$ and 8{}$^\circ$,
respectively.

The beam-width is sufficiently small that the aspect
condition~(\ref{LAUNCH-eq-3}) restricts the number of beam
trajectories to the satellite, for which it is possible to detect
reliably the SAW response in the presence of noise (near the
angles $\gamma=90^\circ$ ). On the other hand,
formula~(\ref{LAUNCH-fig-3}) can be used to determine the
elevation $\theta$ of the wave vector $\boldmath K_t$ of the
shock wave at the known value of the azimuth $\alpha$
(Afraimovich et al., 1998). Hence the phase velocity modulus
$V_t$ can be defined as

\begin{equation}
\label{LAUNCH-eq 8}
V_t=V_h\times sec(\theta)
\end{equation}

The above values of the width $M(\gamma)$ determine the error of
calculation of the elevations $\theta$ (of order 20{}$^\circ$ to the
above conditions (Section~3.3).

\subsection{Determining the position and `switch-on' time of the SAW
source without regard for refraction corrections}
\label{LAUNCH-sect-3.3}

The ionospheric region that is responsible for the main
contribution to TEC variations lies in the neighborhood of the
maximum of the ionospheric $F$-region, which does determine the
height $h_{max}$ of the subionospheric point. When selecting
$h_{max}$, it should be taken into consideration that the
decrease in electron density with height above the main maximum
of the $F_2$-layer proceeds much more slowly than is the case
below the maximum. Since the density distribution with height is
essentially a `weight function' of the TEC response to a wave
disturbance (Afraimovich et al., 1992), it is appropriate to use,
as $h_{max}$, the value exceeding the true height of the layer
$h_{F2}$ maximum by about 100~km. $h_{F2}$ varies over a
reasonably wide range (250--350~km) depending on the time of day
and on some geophysical factors which, when necessary, can be
taken into account if corresponding additional experimental data
and current ionospheric models are used. In all calculations that
follow, $h_{max}=400$~km is used.

To a first approximation, it can be assumed that it is at this
altitude where the imaginary detector is located, which records
the ionospheric SAW response in TEC variations. The `horizontal
size' of the detection region, which can be inferred from the
propagation velocity of the subionospheric point as a consequence
of the orbital motion of the GPS satellite (of order 50~m/s), and
from the SAW period (of order 300~s~--- see
Section~4), does not exceed 15--20~km, which is
far smaller than the `vertical size'.

From the GPS data we can determine the coordinates $X_s$ and
$Y_s$ of the subionospheric point in the horizontal plane $X0Y$
of a topocentric frame of reference centered on the point B(0,0)
at the time of a maximum TEC deviation caused by the arrival of
the SAW at this point (see Fig.~3a). Since we
know the angular coordinates $\theta$ and $\alpha$ of the wave
vector $\boldmath K_t$, it is possible to determine the location
of the point at which this vector intersects the horizontal plane
$X'0Y'$ at the height $h_w$ of the assumed source. Assuming a
rectilinear propagation of the SAW from the source to the
subionospheric point and neglecting the sphericity the coordinate
$X_w$ and $Y_w$ of the source in a topocentric frame of reference
can be defined as

\begin{equation}
\label{LAUNCH-eq-9}
X_w=X_p-\left(h_{max}-h_w\right)
             \frac{\cos\theta\sin\alpha} {\sin\theta}
\end{equation}
\begin{equation}
\label{LAUNCH-eq-10}
Y_w=Y_p-\left(h_{max}-h_w\right)
             \frac{\cos\theta\cos\alpha} {\sin\theta}
\end{equation}

The coordinates $X_w$ and $Y_w$, thus obtained, can readily be
recalculated to the values of the latitude and longitude
($\phi_w$ and $\lambda_w$) of the source.

For SAW generated during earthquakes, industrial explosions and
underground tests of nuclear devices, $h_w$ is taken to be equal
to 0 (the source lying at the ground level). When recording SAW
produced by launchings of powerful rockets, the region of SAW
generation can lie at heights $h_w$ of order 100~km or higher
(Li et al., 1994; Nagorsky, 1998).

Using this approximation we neglect the possible refraction at
the SAW propagation from the source to the height $h_{max}$. In
some publications (Calais et al., 1998) this problem is solved by
performing trajectory calculations using standard `$ray\,\,
tracing$' procedures and neutral atmosphere models. In doing so,
`$ray\,\, tracing$' calculations were carried out from the
source. In this study it is also possible to perform such
calculations, not from the source but from the subionospheric
point (return trajectory).

It should be noted, however, that according to our data
(Section~4) and to the conclusions of some
authors (Li et al., 1994; Nagorsky, 1998), the source lies at
100~km altitude at least. Thus the region of SAW generation is
within the narrow height range (90--150~km) where the sound
velocity gradient is maximal (Li et al., 1994). Within the
approximation of `short' waves (geometrical optics), this
condition would mean the significance of refraction effects.
However, the radial size of the region of SAW formation, like the
extent of the disturbance itself which it generated by the rocket
jet, is about 30--50~km at the above-mentioned heights. This value
is comparable with the typical scale of variation of the sound
velocity with height, hence the wave is no longer a `short' one,
and the geometrical optics approximation is inapplicable. One
would therefore expect that the refraction effect in the
neighborhood of the SAW source will be smaller than anticipated
in terms of geometrical optics.

Given the coordinates of the subionospheric point and of the
disturbance source, the mean value of the SAW propagation
velocity between the source and the subionospheric point, and the
arrival time of the SAW at this point, then within the
approximation of a rectilinear propagation it is easy to
determine the `switch-on' delay $\Delta t_w$ of the anticipated
SAW source with respect to the start. This would make it possible
to obtain additional information about the SAW source which is
needed to understand the mechanism of SAW generation. The
estimates of $\Delta t_w$ made below assume that the propagation
velocity is taken equal to 700~m/s (see Li et al., 1994). Note
that by the `switch-on' time of the source is meant here the time
of a maximum disturbance of the background state of the medium
when the SAW is generated.

\section{Results of measurements}
\label{LAUNCH-sect-4}

Hence, using the transformations described in
Section~3, we obtain the following parameters
determined from TEC variations and characterizing the SAW:
$t_0$~--- start time; $t_p$~--- time of a maximum deviation of
the frequency Doppler shift $F(t)$; $\Delta t$~--- delay of $t_p$
with respect to $t_0$; $T$~--- SAW period; $A_I$~--- TEC
disturbance amplitude; $A_F$~--- amplitude of a maximum frequency
Doppler shift at the `reduced' frequency of 136~MHz; $\alpha $
and $\theta$~--- azimuth and elevation of the wave vector
$\boldmath K_t$; $V_h$ and $V_t$~--- horizontal component and
modulus of the phase velocity; $V_a$~--- mean wave velocity
calculated from the delay $\Delta t$ and from the known
path-length between the launch pad and the subionospheric point;
$\phi_w$ and $\lambda_w$~--- latitude and longitude of the source
at 100km altitude; and $\Delta t_w$~--- `switch-on' delay of the
assumed SAW source with respect to the start.

Corresponding values of the SAW parameters, and also site names
of the GPS-array and GPS satellite PRN numbers are presented in
Table~3 for Baikonur Cosmodrome and Table~4 for the KSC Cosmodrome.

It should be noted that the estimates of $A_I$ and $A_F$ are
obtained by filtering `oblique'\, TEC series; therefore, equivalent
estimates for `vertical'\, TEC are smaller by a factor varying from
1 to 2 depending on the elevation $\theta_s$ of the beam to the
satellite.

Solid curves in Figs.~1 and~2 show trajectories of subionospheric points for each of the GPS
satellites at the height $h_{max}=400$~km. Dark diamonds along
the trajectories correspond to the coordinates of subionospheric
points at the time $t_p$ of a maximum deviation of the frequency
Doppler shift $F$ (Fig.~4c). Asterisks designate
the source location at 100~km altitude inferred from the
GPS-array data. Numbers at the asterisks refer to the
corresponding day numbers and to the `switch-on' delay of the
source with respect to the start time. Dashed straight lines
connecting the anticipated source with the subionospheric point
show the horizontal projection of the wave vector $\boldmath K_t$
(see Fig.~3a).

\subsection{SAW parameters during Proton LV launches from Baikonur
Cosmodrome}
\label{LAUNCH-sect-4.1}

Let us consider the results derived from analyzing the
ionospheric effect of SAW during failed Proton LV launch on
July~5, 1999 (line~2 in Table~2) obtained at
the array (SELE, CHUM,SHAS) for PRN14 (at the left of
Fig.~4, and line~1 in Table~3).

In this case the delay of the SAW response with respect to the
start time is 12~min. The SAW has the form of an $N$-wave with a
period $T$ of about 300~s and an amplitude $A_I=0.5~TECU$, which
is an order of magnitude larger than TEC fluctuations for
background days. It should be noted, however, that this time
interval was characterized by a very low level of geomagnetic
activity (11~nT).

The amplitude of a maximum frequency Doppler shift $A_F$ at the
`reduced' frequency of 136~MHz was found to be 0.12~Hz. In view
of the fact that the shift $F$ is inversely proportional to the
sounding frequency squared (Davies, 1969), this corresponds to a
Doppler shift at the working frequency of 13.6~MHz and the
equivalent oblique-incidence sounding path of about $A_F=12$~Hz.

The azimuth and elevation $\alpha$ and $\theta$ of the wave
vector $\boldmath K_t$ whose horizontal projection is shown in
Fig.~1 by a dashed line and is marked by
$\boldmath K_1$, are 153{}$^\circ$ and 59{}$^\circ$, respectively.
The horizontal component and the modulus of the phase velocity
were found to be $V_h=1808$~m/s and $V_t=931$~m/s. The source
coordinates at 100~km altitude were determined as
$\phi_w=48^\circ$ and $\lambda_w=66^\circ$.

The `switch-on' delay of the SAW source $\Delta t_w$ with respect
to the start time was 264~s. The `mean' velocity of about
$V_a=1000$~m/s, determined in a usual manner from the response
delay with respect to the start, was close the phase velocity
$V_t$.

Similar results for the array (POL2, CHUM, SHAS) and PRN09 were
also obtained for failed launch of October~27, 1999. They
correspond to the projection of the vector $\boldmath K_2$ in
Fig.~1, to the time dependencies in Fig.~4 (at the right), and to line~7 in
Table~3. One can only note that the SAW amplitude was by a factor of 4--5
smaller than that for launch of July~5, 1999. At an increased level of
magnetic activity (--80~nT) this led to a smaller (compared to July~5, 1999)
signal/noise ratio, which, however, did not interfere with obtaining
reliable estimates of SAW parameters.

Fig.~5 (at the right) and line~12 of Table~3 present data for standard
launch of the Proton LV on November~20, 1998. A corresponding projection of
the vector $\boldmath K_3$ is shown in Fig.~1. A
comparison of the data for standard launch November~20, 1998 and
failed launches showed that SAW parameters were reasonably
similar, irrespective of the level of geomagnetic disturbance,
the season, and the local time.

\subsection{SAW parameters during Space Shuttle launches from
the KSC Cosmodrome}
\label{LAUNCH-sect-4.2}

Let us consider the results derived from analyzing the
ionospheric effect of the SAW during Space Shuttle LV launch on
October~29, 1998 (line~3 of Table~2) obtained
at the array (AOML, KYW1, EKY1) for PRN01 (at the left of
Fig.~5, and line~1 in Table~4).

As in the case of Proton LV launches, the delay of the SAW
response with respect to the start time is 12~min. The SAW has
the form of an $N$-wave with a period $T$ of about 210~s and an
amplitude $A_I=0.3\,~TECU$ (for AOML), which is an order of
magnitude larger than TEC fluctuations for background fields. The
amplitude of a maximum frequency Doppler shift $A_F$ at the
`reduced' frequency of 136~MHz was found to be 0.07~Hz (for
AOML). It should be noted, however, that this time interval was
characterized by a very low level of geomagnetic activity
(--15~nT).

The azimuth and elevation $\alpha$ and $\theta$ of the wave
vector $\boldmath K_t$ whose horizontal projection is shown in
Fig.~2 by a dashed line and is marked by
$\boldmath K_1$ are 214{}$^\circ$ and 34.7{}$^\circ$,
respectively. The horizontal component and the modulus of the
phase velocity, and the mean velocity $V_a$ are similar for those
of the Proton~LV (see Table~4). The source
coordinates at 100~km altitude are determined as
$\phi_w=28.5^\circ$ and $\lambda_w=284^\circ$. The `switch-on'
delay of the SAW source $\Delta t_w$ with respect to the start
time is 204~s.

\subsection{Verifying the reliability of measurements of SAW
parameters}
\label{LAUNCH-sect-4.3}

To convince ourselves that the determination of the main
parameters of the SAW form and dynamics is reliable for the
launches analyzed here, in the area of the Baikonur and KSC
Cosmodromes we selected different combinations of three sites out
of the sets of GPS stations available to us, and these data were
processed with the same processing parameters. Relevant results
for Baikonur (including the average results for the sets
$\Sigma$), presented in Table~3 and in
Fig.~1 (SAW source position), show that the
values of SAW parameters are similar, which indicates a good
stability of the data obtained, irrespective of the GPS-array
configuration.

The relative position of the Proton LV flight path and of the
GPS-array stations suitable for SAW detection was very convenient
for our experiment because the subionospheric points of the GPS
satellites were close to the portion of the trajectory where the
anticipated SAW source was located (Li et al., 1994; Nagorsky,
1998), and away from the launch pad. In addition, the aspect
condition (\ref{LAUNCH-eq-3}) corresponding to a maximum
amplitude of the SAW response to the SAW passage was satisfied
quite well for this geometry for all array stations
simultaneously. This is confirmed by a high degree of correlation
of SAW responses at the array elements
(Fig.~4c), which made it possible to obtain
different sets of triangles from the six GPS stations available
to us.

Unlike he Baikonur Cosmodrome, the relative position of the Space
Shuttle LV flight path and of the GPS-array stations was
inconvenient for SAW detection because the subionospheric points
of the GPS satellites were near the launch pad. As a consequence
the aspect condition (\ref{LAUNCH-eq-3}) was not satisfied
simultaneously for all GPS stations located near the Cosmodrome.
This is especially true for stations CCV1, CCV3 and EKY1 located
in the neighborhood of the launch pad along a direction similar
to that of the LV flight path. This resulted in a low degree of
correlation and a very different amplitude of SAW responses at
the array elements (Fig.~5f). Some of the
stations listed in Table~1 are too close to
each other (CCV1, CCV3; MIA1, MIA3; KYW1, KYW2) and cannot be
used in mutual verification of the reliability of measurements.
As a result, we were able to obtain, for the October~29, 1998
launch, only two sets of triangles out of the nine GPS stations
available to us (see Table~4).

Because of the low correlation of the responses and the
inconvenient GPS-array geometry for the April 17, 1998 launch, we
were also unable to obtain reliable estimates of the SAW wave
vector parameters. Therefore, we did not plot the relevant data
in Fig.~2 and limited ourselves to mentioning
them in Table~4 (lines 4 and 5).

We availed ourselves also of another method to check the data
from the GPS-array for reliability, namely, a modeling using the
algorithm for calculating TEC along the `receiver-satellite' beam
for a typical model of the regular ionosphere, a disturbed SAW
with specified properties described by Afraimovich et al. (1998).
A peculiarity of this algorithm is that it is possible to
calculate TEC for a particular selected array and for a real
trajectory of the satellite determined by the initial
navigational RINEX-file.

As an example, Fig.~5 (left) shows the computed
data for the array (SELE, CHUM, SHAS) and for the SAW model in
the form of wave packet (\ref{LAUNCH-eq-5}) with a period of
240~s, the packet's duration of 240~s, a maximum amplitude of
0.5~$TECU$, the horizontal phase velocity $V_h=1200$~m/s, and the
azimuth $\alpha$ and the elevation $\theta$ of the wave vector
$\boldmath K_t$ equal to 153{}$^\circ$ and 60{}$^\circ$,
respectively. A comparison of the parameters specified in the
modeling procedure with those obtained by processing the data of
TEC $I(t)$ model series shows that these values agree with an
accuracy no worse than 10\%. Noteworthy also is a good agreement
of experimental (at the left of Fig.~4) and
model (at the left of Fig.~5) TEC variations
$\Delta I(t)$ and $F(t)$ for the array (SELE, CHUM, SHAS).

\section{Discussion}
\label{LAUNCH-sect-5}

Here we discuss the main results and compare them with findings
reported by other authors. Within the context of this study we
deliberately pass over a physical interpretation of our obtained
results because of no information available about the dynamics
and energetics of Proton and Space Shuttle LV launches. Our
intention is to obtain more trustworthy and reliable data
ensuring the new possibilities of a global GPS monitoring.

\subsection{The form of response}
\label{LAUNCH-sect-5.1}

It was found that in spite of the difference of LV
characteristics, the local time, the season, and the level of
geomagnetic disturbance, the ionospheric response for all
launches has the character of an $N$-wave. The SAW period $T$ is
270--360~s, and the amplitude exceeds the standard deviation of
background TEC fluctuations in this range of periods under quiet
and moderate geomagnetic conditions, by a factor of 2--5 as a
minimum.

Our measurements of the period and amplitude of the SAW response
are in good agreement with frequency Doppler shifts measured in
the HF range during Space Shuttle LV launches on February~28,
1990 and April~28, 1991 (Jacobson and Carlos, 1994), as well as
wit corresponding estimates of a maximum shift $F$ reported by
Nagorsky (1998, 1999) for oblique HF radio path during LV
launches from Baikonur Cosmodrome. They are also similar to the
estimates obtained by Li et al. (1994) using the transionospheric
VHF radio signal from geostationary satellite MARECS-B2 during
Space Shuttle LV launches on October~18, 1993 (STS-58) and
February~3, 1994 (STS-60).

According to measurements reported by Calais and Minster (1996),
during Space Shuttle (STS-58) launch on October~18, 1993 two
series of TEC fluctuations were recorded, one of which had also
the form of an $N$-wave with a maximum amplitude of 0.25~TECU,
which is also consistent with our data.

It was pointed out in the literature that SAW of a similar form
and with a similar amplitude were detected during powerful
industrial explosions (Afraimovich et al., 1984; Jacobson et al.,
1988; Blanc and Jacobson, 1989; Fitzgerald, 1997; Calais et al.,
1998a; 1998b; Nagorsky, 1998).

\subsection{Angular characteristics of the wave vector, and the
phase velocity of SAW}
\label{LAUNCH-sect-5.2}

As pointed out in the Introduction, some researchers report
markedly different values of the SAW propagation velocity~--- up to
several thousand~m/s, which exceeds the sound velocity at SAW
propagation heights in the atmosphere. According to the data from
a review by Karlov et al. (1980), the propagation velocity of the
SAW ionospheric response recorded during launches of the Apollo
mission rockets, varied from 600 to 1670~m/s.

Arendt (1971) suggested that the shock wave produced by rocket
flight is divided in the ionosphere at about 160~km altitude into
the ion-acoustic mode (with a velocity as high as 1.3~km/s) and a
normal acoustic mode (with the velocity of up to 500~m/s). Arendt
(1971) explains the difference in propagation velocities of the
first and second disturbances observed during Apollo-14 and
Apollo-15 launches at the same distance of 1440~km from the
start site, by a difference in atmospheric conditions and
propagation paths of the waves because of seasonal variations of
the ionosphere.

Noble (1990) describes the observations of long-period waves over
the Arecibo incoherent scatter station during Space Shuttle
STS-4 launch on June~27, 1982. At a large distance from the path
(up to 1000~km) the group velocity of wave propagation was found
to be 600--700~m/s.

According to GPS measurements (Calais and Minster, 1996), the SAW
phase velocity at ionospheric heights is about 1000--1300~m/s.

A common limitation of existing methods for determining the SAW
phase velocity is the need to know the launch time of the rocket
since this velocity is calculated from the SAW delay with respect
to the start time assuming that the velocity along the
propagation path is constant, which is by far contrary to fact.
Furthermore, only the horizontal component of the phase velocity
$V_h$ was in essence determined in such studies. At different
values of elevation of the wave vector $\boldmath K_t$, the
velocity $V_h$ corresponds to markedly differing values of the
modules of the phase velocity $V_t$.

The use of the method proposed in this paper makes possible
determining angular characteristics of the wave vector $\boldmath
K_t$ and, accordingly, estimating $V_t$. According to our data
(Tables~3 and~4), the
elevation of the SAW wave vector varied from 30--60${}^\circ$,
and the SAW phase velocity was in the range of from 900 to
1200~m/s. We determine the phase velocity of the line of equal
TEC at the height of the ionospheric $F$-region maximum which
makes the main contribution to variations in TEC between the
receiver and the GPS satellite and corresponds to the region of
maximum sensitivity of the method. Since $V_t$ approaches the
sound velocity at these altitudes (Li et al., 1994), this makes
it possible to identify the sound nature of a TEC perturbation.

\subsection{The location and `switch-on' delay of the SAW source}
\label{LAUNCH-sect-5.3}

The position of the SAW source, calculated by neglecting
refraction corrections, corresponds to the segment of the LV
trajectory (Figs.~1 and~2)
within distances of at least 700--900~km from the launch pad for
the Proton LV and of at least 200--500~km for the Space
Shuttle LV. This is consistent with the `switch-on' delay
$\Delta t_w$ of the source which is 250--300~s for the Proton LV
and 200~s for the Space Shuttle LV. As is evident from our
data, the calculated position of the SAW source for rocket
launches does not coincide with the position of the launch pad.
At the same time the source location is in reasonably good
agreement with that of horizontal projections of LV trajectories
(Figs.~1 and~2).

Kaschak et al. (1970) analyzed the data from the infra sound
measuring arrays located on the USA north-eastern coast which
observed strong acoustic signals from launch and reentry areas of
the Saturn-5 LV. The authors identified three types of signals:
one includes early signals whose arrival time corresponded to
supersonic values of their velocity equal to 500--1000~m/s; the
other type represents normal signals with the group velocity
approximately equal to the normal velocity of sound propagation
in the air; and the last type involves late signals with subsonic
velocities in the range of from 190 to 240~m/s. Kaschak et al.
(1970) and Balachandran et al. (1971) suggested that the
so-called `early' signals during rocket launches are caused by
SAW which are generated during the reentry of the first stage at
distances from the launch pad exceeding 500~km.

However, our data are in better agreement with the mechanism
substantiated by Li et al. (1994), Calais and Minster (1996), and
Nagorsky (1998, 1999). They believe that the generation of SAW
occurs during a nearly horizontal travel of the rocket with the
operating engine along the acceleration segment of the
trajectory, at the lower atmospheric heights of 100--130~km. The
rocket travels this segment with supersonic velocity at 100--300~s
of its flight at a distance of at least 500~km from the launch
pad (see the data on the Proton LV flight schedule in
Section~2). As soon as it ascends to an
altitude of about 100~km, the SAW source is `switched on'.

\section{Conclusions}
\label{LAUNCH-sect-6}

In this paper we have investigated the form and dynamics of
shock-acoustic waves (SAW) generated during rocket launches. We
have developed a method for determining SAW parameters (including
angular characteristics of the wave vector and the SAW phase
velocity, as well as the direction to the source) using the
GPS-arrays whose elements can be chosen out of a large set of GPS
stations forming part of a global IGS network. Unlike existing
radio techniques, the proposed method estimates SAW parameters
without a priory information about the site and time of rocket
launch. The implementation of the method is illustrated by
analyzing ionospheric effects from launches of the Proton and
Space Shuttle LV from Baikonur and KSC Cosmodromes in 1998 and
1999 (totaling five launches).

The results reported in this study suggest the following
conclusions:

\begin{enumerate}

\item
In spite of the difference of LV characteristics, for all
launches the ionospheric response has the character of an
$N$-wave corresponding to the form of a shock wave, irrespective
of the type of disturbance source (rocket launch, industrial
explosion).

\item
The SAW period $T$ is 270--360~s, and its amplitude (from 0.1 to
0.5~$TECU$) exceeds the standard deviation of $TECU$ background
fluctuations in this range of periods under quiet and moderate
geomagnetic conditions by a factor of 2--5 as a minimum.

\item
The elevation of the SAW wave vector varies within
35--60${}^\circ$, and the SAW phase velocity (900--1200~m/s)
approaches the sound velocity at heights of the ionospheric
$F$-region minimum. This makes it possible to identify the sound
nature of a TEC perturbation.

\item
The position of the SAW source, calculated by neglecting
refraction corrections, corresponds to a segment of the Proton
LV trajectory at a distance of at least 700--900~km from the
launch pad and to the LV flight altitude of at least 100~km. For
the Space Shuttle LV the position of the SAW source corresponds
to the segment of the trajectory at a distance of at least
200--500~km from the launch pad and to the LV flight altitude of
at least 100~km.

\end{enumerate}

Hopefully, our investigation would provide additional useful
insights into the physical processes occurring during flights of
rockets in the Earth's atmosphere on the acceleration segment of
the trajectory. In addition, this would make it possible to
identify more reliable signal indications of technogenic effects
which are necessary for constructing an effective global radio
subsystem for detection and localization of these effects by
processing the data from an international network of
two-frequency receivers of the GPS-GLONASS navigation systems.

\section{Acknowledgements}
\label{LAUNCH-sect-7}

We are grateful to E.~A.~Ponomarev, V.~V.~Yevstafiev,
A.~M.~Uralov, P.~M.~Nagorsky, N.~N.~Klimov, and A.~D.~Kalikhman
for their interest in this study, helpful advice and active
participation in discussions. Authors are grateful to
~K.~S.~Palamartchouk and ~O.~S. Lesuta for preparing the input
data. Thanks are also due to V.~G.~Mikhalkosky for his assistance
in preparing the English version of the manuscript. This work was
done with support from the Russian foundation for Basic Research
(grants 97-02-96060 and 99-05-64753), as well as under grant 1999
of the RF Ministry of Education (Minvuz), under direction of
B.~O.~Vugmeister.

{}

\begin{thebibliography}{99}

\bibitem{Afr84}
Afraimovich,~E.L., Varshavsky,~I.I., Vugmeister,~B.O., et~al.,
1984. Influence of surface industrial explosions on Doppler and
angular characteristics of the ionosphere-reflected radio signal.
Geomagnetizm i aeronomiya, 24, 322--324 (in Russian).

\bibitem{Afr92}
Afraimovich,~E.L., Terechov,~A.I., Udodov,~M.Yu., and
Fridman,~S.V., 1992. Refraction distortions of
trans\-iono\-sphe\-ric radio signals caused by changes in a
regular ionosphere and by travelling ionospheric disturbances.
Journal of Atmospheric and Solar-Terrestrial Physics, 54,
1013--1020.

\bibitem{Afr98}
Afraimovich,~E.L., Palamartchouk,~K.S., and Perevalova,~N.P.,
1998. GPS radio interferometry of travelling ionospheric
disturbances. Journal of Atmospheric and Solar-Terrestrial
Physics, 60, 1205--1223.

\bibitem{Are71}
Arendt,~P.R., 1971. Ionospheric undulations following
`Appolo--14' launching. Nature, 231, 438--439.

\bibitem{Bal71}
Balachandran,~N.K., Donn,~W.L., and Kaschak,~G., 1971. On the
propagation of infrasound from rockets: effects of winds. Journal
of the Acoustical Society of America, 50 (2), part~1, 397--404.

\bibitem{Ber76}
Bertel,~L., Bertin,~F., and Testud,~J., 1976. De la mesure du
contenu electronique integre appliquee a l'observation des ondes
de gravite de moyenne echelle. Journal of Atmospheric and
Solar-Terrestrial Physics, 38, 261--270.

\bibitem{Bla89}
Blanc,~E., and Jacobson,~A.R., 1989. Observation of ionospheric
disturbances following a 5-kt chemical explosion. 2.~Prolonged
anomalies and stratifications in the lower thermosphere after
shock passage. Radio Science, 24, 739--746.

\bibitem{Cal95}
Calais,~E. and Minster,~J.B., 1995. GPS detection of ionospheric
perturbations following the January 1994, Northridge earthquake.
Geophysical Research Letters, 22, 1045--1048.

\bibitem{Cal96}
Calais,~E., and Minster,~J.B., 1996. GPS detection of ionospheric
perturbations following a Space Shuttle ascent. Geophysical
Research Letters, 23, 1897--1900.

\bibitem{Cal98a}
Calais,~E., Minster,~J.B., Hofton,~M.A., and Hedlin,~M.A.H.,
1998a. Ionospheric signature of surface mine blasts from Global
Positioning System measurements. Geophysical Journal
International, 132, 191--202.

\bibitem{Cal98b}
Calais,~E., Minster,~J.B., and Bernard,~J., 1998b. GPS,
Earthquake, the ionosphere and Space Shuttle. Physics of Earth
and Planet, 105, 167--181.

\bibitem{Dav69}
Davies,~K., 1969. Ionospheric radio waves. Blaisdell Publishing
Company. A Division of Ginn and Company. Waltham,
Massachusetts-Toronto-London.

\bibitem{Fit97}
Fitzgerald,~T.J., 1997. Observations of total electron content
perturbations on GPS signals caused by a ground level explosion.
Journal of Atmospheric and Solar-Terrestrial Physics, 59,
829--834.

\bibitem{Gur93}
Gurtner,~W., 1993. RINEX:~The Receiver Independent Exchange
Format Version~2.
http://igscb.jpl.nasa.gov/igscb/data/format/rinex2.txt

\bibitem{Hof92}
Hofmann-Wellenhof,~B., Lichtenegger,~H., and Collins,~J., 1992.
Global Positioning System: Theory and Practice, Springer-Verlag
Wien, New York.

\bibitem{Jac88}
Jacobson,~A.R., Carlos,~R.C., and Blanc,~E., 1988. Observation of
ionospheric disturbances following a 5-kt chemical explosion.
1.~Persistent oscillation in the lower thermosphere after shock
passage. Radio Science, 23, 820--830.

\bibitem{Jac94}
Jacobson,~A.R., and Carlos,~R.C., 1994. Observations of
acoustic--gravity waves in the thermosphere following Space
Shuttle ascents. Journal of Atmospheric and Solar-Terrestrial
Physics, 56, 525--528.

\bibitem{Kar80}
Karlov,~V.D., Kozlov,~S.I., and Tkachev,~G.N., 1980. Large-scale
disturbances in the ionosphere produced by rocket flight with the
operating engine. Kosmicheskiye issledovaniya, 18, 266--277 (in
Russian).

\bibitem{Kas70}
Kaschak,~G, Donn,~W.L., and Fehr,~U., 1970. Long-range infrasound
from rockets. Journal of the Acoustical Society of America, 48
(1), part~1, 12--20.

\bibitem{Klo97}
Klobuchar,~J.A., 1997. Real-time ionospheric science: The new
reality. Radio Science, 32, 1943--1952.

\bibitem{Li94}
Li,~Y.Q., Jacobson,~A.R., Carlos,~R.C., Massey,~R.S.,
Taranenko,~Y.N., and Wu,~G., 1994. The blast wave of the Shuttle
plume at ionospheric heights. Geophysical Research Letters, 21,
2737--2740.

\bibitem{Men81}
Mendillo,~M., 1981. The effects of rocket launches of the
ionosphere. Advances in Space Research, 1, 275--290.

\bibitem{Men82}
Mendillo,~M., 1982. Modification of the ionosphere by large space
vehicles. Advances of Space Research, 2, 150--159.

\bibitem{Mer97}
Mercier,~C. and Jacobson,~A.R., 1997. Observations of atmospheric
gravity waves by radio interferometry: are results biased by the
observational technique? Annales G\'eophysicae, 15,
430--442.

\bibitem{Nag98}
Nagorsky,~P.M., 1998. The inhomogeneous structure of the
ionospheric F-region produced by rockets. Geomagnetizm i
aeronomiya, 38, 100--106 (in Russian).

\bibitem{Nag99}
Nagorsky,~P.M., 1999. Analysis of the HF radio signal response
to ionospehric plasma disturbances caused by shock-acoustic
waves. Izvestya VUZov Radiofizika, 42, 36--44 (in Russian).

\bibitem{Nob90}
Noble,~S.T., 1990. A large-amplitude traveling ionospheric
disturbance exited by the Space Shuttle during launch. Journal of
Geophysical Research, 95, 19,037--19,044.

\end{thebibliography}
\end{document}